\documentclass[10pt, conference, letterpaper]{IEEEtran}
\IEEEoverridecommandlockouts    

\usepackage{multirow}
\usepackage{amsmath}
\usepackage{amssymb}
\usepackage{graphicx}
\graphicspath{{./Figures/}}
\usepackage{booktabs}
\usepackage{array}
\usepackage{xcolor}
\usepackage[hidelinks]{hyperref}
\usepackage{balance}
\usepackage{tikz}
\usetikzlibrary{arrows.meta,positioning,calc,fit}

\usepackage{geometry}
\usepackage{afterpage}

\afterpage{
  \newgeometry{
    top=0.75in,
    bottom=0.75in,
    left=0.75in,
    right=0.75in
  }
}

\title{\LARGE \bf From GPU to Microcontroller: Online Ridge Regression\\for Edge-Deployable Traffic Prediction}

\author{
\IEEEauthorblockN{Suresh Purini, Archit Narwadkar, and Deepak Gangadharan}
\IEEEauthorblockA{Computer Systems Group\\
International Institute of Information Technology, Hyderabad, India\\
\{suresh.purini, deepak.g\}@iiit.ac.in, archit.narwadkar@research.iiit.ac.in}
}

\begin{document}

	\maketitle
	\thispagestyle{empty}
	\pagestyle{empty}

    \begin{abstract}
State-of-the-art traffic flow forecasting models, including Graph Convolutional Networks and graph-less MLPs, require centralized GPU training across all sensors, making them impractical for resource-constrained intelligent transportation deployments.
We show that much of this complexity is unnecessary.
A parametric analysis of the recent graph-less model GLMST~\cite{zhang2025glmst} reveals that reducing its internal embedding dimension from 64 to 4 degrades MAPE by less than one percentage point, suggesting that the model's effective capacity far exceeds what the task requires.
Motivated by this finding, we replace the neural architecture entirely with per-sensor Ridge regression using horizon-aligned periodic features, combined with Recursive Least Squares (RLS) for online adaptation.
With only 444~parameters per sensor ($80{\times}$ fewer than GLMST) and test-time online adaptation, our method achieves the best MAPE on three of four PEMS benchmarks, and remains within one percentage point on the fourth.
Because each sensor's model is self-contained and involves only elementary linear algebra, the entire pipeline (training, inference, and online adaptation) runs on edge hardware without a GPU.
An ESP32 microcontroller (160~MHz, 520~KB SRAM) completes cold-start training in 7.4~s and each predict-and-update in under 2~ms with zero heap allocation; a single Raspberry Pi~5 core completes cold-start training in 0.21~s and each predict-and-update in 0.26~ms.
\end{abstract}

    \section{Introduction}
\label{sec:introduction}

Short-term traffic flow forecasting is a core component of Intelligent Transportation Systems (ITS), enabling real-time traffic signal control, route guidance, and congestion management.
Given historical sensor readings, the task is to predict traffic flow at each sensor for the next hour (12 steps at 5-minute intervals).
The dominant approach uses Graph Convolutional Networks (GCNs) that jointly model spatial dependencies across sensors and temporal dynamics within each sensor~\cite{yu2018stgcn,li2018dcrnn,wu2019graphwavenet,bai2020agcrn,choi2022stgncde}.
While these models have steadily improved accuracy on standard benchmarks, each generation introduces additional architectural complexity, such as attention mechanisms, adaptive adjacency matrices, and neural differential equations, with diminishing marginal gains.

Beyond accuracy, these models share a practical constraint: they require collecting sensor data at a central server and training jointly on a GPU.
For large-scale ITS deployments spanning hundreds of sensors, this imposes significant data movement, compute infrastructure, and maintenance costs, with no path to on-device training or adaptation.

Recently, GLMST~\cite{zhang2025glmst} challenged the necessity of graph convolutions by showing that a pure-MLP architecture with per-node affine parameters achieves competitive accuracy.
However, GLMST still trains all sensors jointly on a GPU, leaving the centralized infrastructure requirement unchanged.
This raised two questions: how much of the neural complexity is actually needed, and can the centralized training requirement be eliminated entirely?

In this paper, we push this question to its logical conclusion.
We first analyze GLMST by systematically reducing its internal dimensions: the embedding width~$e$, the per-node layer width~$s$, and the network depth~$L$.
A grid search reveals that reducing $e$ from 64 to~4 and $s$ from 64 to~24, a $16{\times}$ per-sensor parameter reduction, degrades MAPE by less than one percentage point.
This suggests that the model's effective capacity far exceeds what the task requires, since the input is a single scalar (flow) per timestep.

Motivated by this insight, we replace the neural temporal mixer entirely with per-sensor Ridge regression using horizon-aligned periodic features.
For each sensor and forecast horizon, we construct a 36-dimensional feature vector from recent, daily-aligned, and weekly-aligned flow lags, and fit a Ridge model in closed form.
Adding Recursive Least Squares (RLS) online adaptation during the test period, our method achieves the best MAPE on three of four PEMS benchmarks and remains within one percentage point on the fourth, while using only 444~parameters per sensor ($80{\times}$ fewer than GLMST).
Because each sensor's model is independent and requires only simple linear algebra, the entire pipeline (training, inference, and online adaptation) runs on highly resource-constrained edge devices.

Our contributions are:
\begin{enumerate}
\item A parametric analysis of GLMST showing that its per-sensor parameter count can be reduced from 35{,}500 to under 2{,}200 with less than one percentage point MAPE loss, and that inference FLOPs drop from 974K to under 55K, revealing that the model's effective capacity far exceeds what the task requires.
\item A per-sensor Ridge regression model using horizon-aligned periodic features and Recursive Least Squares online adaptation. It matches or exceeds neural baselines on 3/4 PEMS datasets and remains within one percentage point on the fourth, with only 444~parameters and 876~inference FLOPs per sensor. The model requires no centralized training, no GPU, and no inter-sensor data exchange.
\item End-to-end edge deployment validated on an ESP32 microcontroller (160~MHz, 520~KB SRAM) and a Raspberry Pi~5, where each sensor trains, predicts, and adapts using only local data. On the ESP32, cold-start training completes in 7.4~s and each predict-and-update cycle runs in under 2~ms with zero heap allocation within 5~KB of stack. On a single Raspberry Pi~5 core, cold-start training completes in 0.21~s and each predict-and-update cycle runs in 0.26~ms.
\end{enumerate}

Section~\ref{sec:related} reviews related work. Section~\ref{sec:method} presents the GLMST architecture and our parametric analysis revealing its over-parameterization, introduces the horizon-aligned periodic feature design and per-sensor Ridge regression, describes the RLS online adaptation mechanism, and details the distributed edge deployment pipeline. Section~\ref{sec:experiments} reports experimental results and edge deployment benchmarks. Section~\ref{sec:conclusion} concludes.

    \section{Related Work}\label{sec:related}

\textbf{GCN-based traffic forecasting.}
Graph convolutional networks have dominated traffic forecasting since STGCN~\cite{yu2018stgcn} combined graph convolutions with temporal convolutions.
Subsequent work introduced diffusion convolutions (DCRNN~\cite{li2018dcrnn}), adaptive adjacency learning (Graph WaveNet~\cite{wu2019graphwavenet}), attention mechanisms (ASTGCN~\cite{guo2019astgcn}), adaptive graph generation (AGCRN~\cite{bai2020agcrn}), spatial-temporal fusion (STFGNN~\cite{li2021stfgnn}), and neural controlled differential equations (STG-NCDE~\cite{choi2022stgncde}).
Transformer-based approaches have further increased complexity, with PDFormer~\cite{jiang2023pdformer} modeling propagation delays and STAEformer~\cite{liu2023staeformer} using adaptive embeddings.
Notably, ASTGCN uses three independent components to model recent, daily-periodic, and weekly-periodic dependencies, a temporal decomposition conceptually similar to our horizon-aligned features.
However, each component employs graph convolutions and spatial-temporal attention, resulting in significantly higher complexity without matching the accuracy of our linear approach.
Each generation adds parameters and architectural complexity, yet the marginal accuracy gains have diminished.

\textbf{Graph-less methods.}
GLMST~\cite{zhang2025glmst} challenged the necessity of graph convolutions by showing that a pure-MLP architecture with per-node affine parameters in place of graph layers achieves competitive accuracy.
Its design draws on MLP-Mixer~\cite{tolstikhin2021mlpmixer}, which demonstrated that simple channel-wise and token-wise MLPs can match attention-based vision models.
Our work extends this line of inquiry: if graph convolutions are unnecessary, is the MLP itself necessary?

\textbf{Classical and linear methods.}
Statistical methods such as VAR~\cite{lutkepohl2005var} and ARIMA were standard before deep learning but are now treated as weak baselines.
FC-LSTM~\cite{hochreiter1997lstm} bridged the gap to neural approaches.
Ridge regression~\cite{hoerl1970ridge} has seen limited use in traffic forecasting, typically without periodic feature engineering.
We show that with horizon-aligned daily and weekly features, Ridge regression matches or exceeds neural methods on several benchmarks.

\textbf{Online and adaptive methods.}
Recursive Least Squares (RLS)~\cite{haykin2014adaptive} and Kalman filtering are well-established in adaptive signal processing but rarely applied to traffic flow prediction at scale.
Online adaptation is particularly relevant for traffic, where patterns shift due to construction, incidents, and seasonal changes.
Our RLS extension warm-starts from a batch Ridge solution and adapts during the test period, closing the gap to neural methods without retraining.

\textbf{Edge and embedded deployment.}
The TinyML paradigm~\cite{warden2019tinyml} has enabled on-device inference for applications such as keyword spotting and gesture recognition on microcontrollers, but its adoption in traffic forecasting remains limited.
Existing edge computing approaches for traffic prediction typically offload neural network inference to GPU-equipped edge servers rather than running models directly on sensor-attached microcontrollers.
To our knowledge, no prior work has demonstrated end-to-end traffic flow prediction, including both training and online adaptation, on a microcontroller-class device.

	\section{Proposed Method}
\label{sec:method}

\subsection{Problem Formulation}

Consider a road network with $N$ sensors, each recording traffic flow at 5-minute intervals.
Let $x_i(t) \in \mathbb{R}$ denote the flow at sensor $i$ at time $t$.
Given a historical window of $T{=}12$ timesteps (one hour), we predict the next $H{=}12$ timesteps:
\begin{equation}
\hat{x}_i(t{+}h) = f_h\bigl(\mathbf{x}_i(t{-}T{+}1:t)\bigr), \quad h = 1, \ldots, H.
\label{eq:problem}
\end{equation}
Unlike GCN-based methods that jointly model all $N$~sensors using a graph, we train an \emph{independent model per sensor per horizon}, exploiting the observation that temporal patterns dominate spatial ones for flow prediction.

\subsection{Horizon-Aligned Periodic Features}

Traffic flow exhibits strong daily and weekly periodicities.
A na\"ive approach concatenates the same historical lag (e.g., $x_i(t{-}288)$ for the daily lag at 5-min resolution) regardless of the target horizon.
We instead use \emph{horizon-aligned} lags: for horizon~$h$, the daily-aligned feature is $x_i(t{+}h{-}288)$, i.e., the flow observed exactly 24~hours before the \emph{target} timestep.

Concretely, the feature vector for sensor $i$, horizon $h$, at time $t$ is:
\begin{multline}
\mathbf{z}_i^{(h)}(t) = \bigl[\underbrace{x_i(t), \ldots, x_i(t{-}11)}_{\text{12 recent}},\\
\underbrace{x_i(t{+}h{-}288), \ldots}_{\text{12 daily}},\; \underbrace{x_i(t{+}h{-}2016), \ldots}_{\text{12 weekly}}\bigr]
\label{eq:features}
\end{multline}
where the daily-aligned block contains $x_i(t{+}h{-}288{-}j)$ and the weekly-aligned block contains $x_i(t{+}h{-}2016{-}j)$ for $j{=}0,\ldots,11$.
This yields $d{=}36$ features per sample.

Horizon alignment is critical: it ensures that each periodic lag corresponds to the \emph{same time-of-day and day-of-week} as the prediction target, not the input window.
Without this alignment, the periodic features carry a systematic offset of up to one hour, which we find degrades MAPE by 1--2 percentage points.

\subsection{Per-Sensor Ridge Regression}

For each sensor $i$ and horizon $h$, we fit a Ridge regression model~\cite{hoerl1970ridge}:
\begin{equation}
\hat{x}_i(t{+}h) = \mathbf{w}_{i,h}^\top \mathbf{z}_i^{(h)}(t) + b_{i,h}
\label{eq:ridge}
\end{equation}
where $\mathbf{w}_{i,h} \in \mathbb{R}^d$ and $b_{i,h} \in \mathbb{R}$ are obtained by minimizing the regularized least-squares objective:
\begin{equation}
\min_{\mathbf{w}, b} \sum_{t \in \mathcal{T}_\text{train}} \bigl(x_i(t{+}h) - \mathbf{w}^\top \mathbf{z}_i^{(h)}(t) - b\bigr)^2 + \alpha \|\mathbf{w}\|_2^2
\label{eq:ridge_obj}
\end{equation}
with $\alpha{=}1.0$.
Appending the bias into the weight vector $\tilde{\mathbf{w}} = [\mathbf{w}; b]$ and augmenting each feature vector as $\tilde{\mathbf{z}} = [\mathbf{z}; 1]$, the closed-form solution is:
\begin{equation}
\tilde{\mathbf{w}}_{i,h} = \bigl(\tilde{\mathbf{Z}}^\top \tilde{\mathbf{Z}} + \alpha \mathbf{I}\bigr)^{-1} \tilde{\mathbf{Z}}^\top \mathbf{y}
\label{eq:ridge_closed}
\end{equation}
where $\tilde{\mathbf{Z}} \in \mathbb{R}^{n \times (d+1)}$ is the augmented design matrix and $\mathbf{y} \in \mathbb{R}^n$ is the target vector.
This requires only a single matrix inversion of size $(d{+}1){\times}(d{+}1)$, completing in milliseconds per model.
The total model for one sensor comprises $H \times (d{+}1) = 12 \times 37 = 444$ parameters across all horizons.

\subsection{GLMST Architecture and Parametric Analysis}
\label{sec:glmst_bottleneck}

To understand \emph{why} a linear model can match a neural network, we first review the GLMST architecture~\cite{zhang2025glmst} in matrix form and then introduce a parametric variant that reveals its effective dimensionality.

\textbf{GLMST architecture.}
GLMST is a graph-less MLP that replaces graph convolutions with per-node affine transformations.
We describe the forward pass for a single sensor~$n$, parameterized by an \emph{embedding dimension}~$e$ (the hidden size carried between blocks and used in temporal mixing) and a \emph{spatial dimension}~$s$ (the hidden size used for per-node spatial mixing).
In the original GLMST, $e{=}s{=}64$; our parametric variant (below) decouples them.
The input is $\mathbf{X}_n \in \mathbb{R}^{T \times d_\text{in}}$, where $T{=}12$ timesteps and $d_\text{in}{=}1$ (flow only).

\emph{Step~1: Input projection.}
A shared linear layer maps to a hidden representation:
\begin{equation}
\mathbf{H}_n^{(0)} = \mathbf{X}_n \mathbf{W}_\text{in} + \mathbf{1}\mathbf{b}_\text{in}^\top \;\in\; \mathbb{R}^{T \times e}
\label{eq:glmst_input}
\end{equation}
where $\mathbf{W}_\text{in} \in \mathbb{R}^{d_\text{in} \times e}$ and $\mathbf{b}_\text{in} \in \mathbb{R}^{e}$.

\emph{Step~2: Spatio-temporal blocks.}
Each of the $L$ blocks applies two sub-layers with residual connections.
Let $\mathbf{H} = \mathbf{H}_n^{(\ell-1)} \in \mathbb{R}^{T \times e}$ denote the input to block~$\ell$.

The \emph{Cross-Channel Temporal Mixing (CCTM)} sub-layer first applies a feature-wise linear transformation, then mixes across time:
\begin{align}
\mathbf{M} &= \sigma(\mathbf{H}\,\mathbf{W}_{t1} + \mathbf{1}\mathbf{b}_{t1}^\top) & &\text{(feature mixing)} \label{eq:cctm_feat} \\
\text{CCTM}(\mathbf{H}) &= \sigma(\mathbf{W}_{t2}\,\mathbf{M} + \mathbf{b}_{t2}) & &\text{(time mixing)} \label{eq:cctm_time}
\end{align}
where $\mathbf{W}_{t1} \in \mathbb{R}^{e \times e}$ and $\mathbf{b}_{t1} \in \mathbb{R}^e$ operate along features (right-multiplied, shared across timesteps), $\mathbf{W}_{t2} \in \mathbb{R}^{T \times T}$ and $\mathbf{b}_{t2} \in \mathbb{R}^{T}$ mix across the time axis (left-multiplied), and $\sigma$ is ReLU.
With residual connection: $\widetilde{\mathbf{H}} = \mathbf{H} + \text{CCTM}(\mathbf{H})$.

The \emph{Graph-Less Spatial Mixing (GLSM)} sub-layer projects from $e$ to $s$ dimensions, applies LayerNorm and \emph{node-specific} affine parameters that replace graph convolution:
\begin{align}
\mathbf{S} &= \text{LN}(\widetilde{\mathbf{H}}\,\mathbf{W}_{s1} + \mathbf{1}\mathbf{b}_{s1}^\top) & &\text{(shared)} \label{eq:glsm_shared} \\
\text{GLSM}_n(\widetilde{\mathbf{H}}) &= \mathbf{S} \odot \mathbf{1}{\mathbf{w}_{s2}^{(n)}}^\top + \mathbf{1}{\mathbf{b}_{s2}^{(n)}}^\top & &\text{(per-node)} \label{eq:glsm_node}
\end{align}
where $\mathbf{W}_{s1} \in \mathbb{R}^{e \times s}$ and $\mathbf{b}_{s1} \in \mathbb{R}^{s}$ are shared across all sensors, $\text{LN}$ denotes LayerNorm, and $\mathbf{w}_{s2}^{(n)}, \mathbf{b}_{s2}^{(n)} \in \mathbb{R}^{s}$ are the \emph{per-node} affine parameters that give each sensor its own identity.
A second shared linear layer projects the result back to $e$ dimensions before the residual connection: $\mathbf{H}_n^{(\ell)} = \widetilde{\mathbf{H}} + \text{GLSM}_n(\widetilde{\mathbf{H}})\,\mathbf{W}_{s3}$, where $\mathbf{W}_{s3} \in \mathbb{R}^{s \times e}$.

\emph{Step~3: Output projection.}
A linear layer maps back to $d_\text{in}$ dimensions:
$\hat{\mathbf{Y}}_n = \mathbf{H}_n^{(L)} \mathbf{W}_\text{out} + \mathbf{1}\mathbf{b}_\text{out}^\top \in \mathbb{R}^{H \times d_\text{in}}$.
Note that GLMST uses $T{=}H{=}12$, so the output directly produces all $H$~forecast horizons.

\textbf{Parameter breakdown.}
With $e{=}s{=}64$ and $L{=}4$, the shared parameters per block are $\mathbf{W}_{t1} \in \mathbb{R}^{e \times e}$, $\mathbf{W}_{s1} \in \mathbb{R}^{e \times s}$, and $\mathbf{W}_{t2} \in \mathbb{R}^{T \times T}$, plus biases and LayerNorm.
The node-specific parameters are $\mathbf{w}_{s2}^{(n)}, \mathbf{b}_{s2}^{(n)} \in \mathbb{R}^{s}$ per block, totaling $2Ns$ per block and $2NsL$ overall.
On PEMS03 ($N{=}358$), these $2 \times 358 \times 64 \times 4 = 183{,}296$ node-specific parameters constitute \textbf{84\%} of the 218K total.
Since node-specific parameters scale with $Ns$, model size varies across datasets (122K--487K for PEMS08--PEMS07; see Table~\ref{tab:complexity}).

\textbf{Per-sensor cost.}
For inference at a single sensor, GLMST requires all 35K shared parameters plus $2sL{=}512$ node-specific parameters, totaling approximately 35{,}500~parameters per sensor, an $80{\times}$ reduction to Ridge's 444.
However, the total parameter count remains significant because GLMST requires \emph{centralized training}: all $N$~sensors' flow data must be collected at a central server to jointly optimize the shared weights, and the trained model (35K shared $+$ 512 per-node parameters) must then be distributed back to each sensor.
This centralized pipeline contrasts with our approach, where each sensor trains and updates its model entirely from local data.

\textbf{Parametric variant: GLMST($e$, $s$, $L$).}
A structural observation suggests that $e{=}64$ is far larger than necessary.
Since $d_\text{in}{=}1$, the input projection produces $\mathbf{H}^{(0)} = \mathbf{x}_n \mathbf{w}_\text{in}^\top + \mathbf{1}\mathbf{b}_\text{in}^\top$, where the informative term $\mathbf{x}_n \mathbf{w}_\text{in}^\top$ is a \emph{rank-1 matrix} --- every column is a scalar multiple of the same flow vector.
Regardless of the embedding dimension~$e$, the initial hidden state carries only $T{=}12$ scalar values of information.
CCTM's feature mixing ($\mathbf{W}_{t1}$) applied to this rank-1 input yields another rank-1 output (before the nonlinearity), so the temporal representation is inherently low-dimensional.
This suggests that $e$ can be drastically reduced without meaningful information loss.

For the spatial dimension~$s$, the argument is less structural: $\mathbf{w}_{s2}^{(n)} \in \mathbb{R}^s$ acts as each sensor's learned identity, and whether a large $s$ is needed depends on how many distinct traffic patterns exist across sensors.

To test these hypotheses, we introduce a parametric variant that allows $e \ll s$, reducing CCTM capacity while preserving per-node spatial expressiveness.

\textbf{Key insight.}
We perform a grid search over $e \in \{1, 2, 4\}$, $s \in \{8, 16, 24, 32\}$, and $L \in \{2, 3, 4\}$ on PEMS03 (Section~\ref{sec:nas}).
The best configuration, GLMST(4, 24, 4), achieves 15.42\% MAPE with 71K parameters --- only 0.68 percentage points behind the full GLMST (14.74\%) despite a $3{\times}$ parameter reduction.
Even GLMST(4, 8, 2) with just 12K parameters, an $18{\times}$ reduction, reaches 15.73\%, confirming that $s{=}8$ suffices empirically.
These results validate the rank-1 intuition: the temporal mixing operates in a \emph{low-rank subspace of dimension $\leq 4$}, motivating the replacement of the neural temporal mixer with a linear model.

\subsection{RLS Online Adaptation}

Static Ridge regression cannot adapt to distribution shifts during the test period (e.g., changing traffic patterns due to construction or special events).
We address this with Recursive Least Squares (RLS)~\cite{haykin2014adaptive}, which updates the Ridge solution online as each new observation arrives.
The key idea is simple: rather than re-solving the Ridge objective from scratch whenever new data becomes available, RLS makes a small correction to the existing weight vector based on the prediction error of the latest observation.

Concretely, given a new observation $(\mathbf{z}_t, y_t)$ at test time, RLS performs three steps derived from the Sherman--Morrison formula~\cite{sherman1950adjustment}:
\begin{align}
\mathbf{k}_t &= \frac{\mathbf{P}_{t-1} \mathbf{z}_t}{\lambda + \mathbf{z}_t^\top \mathbf{P}_{t-1} \mathbf{z}_t} \label{eq:rls_gain} \\
\mathbf{w}_t &= \mathbf{w}_{t-1} + \mathbf{k}_t (y_t - \mathbf{w}_{t-1}^\top \mathbf{z}_t) \label{eq:rls_weight} \\
\mathbf{P}_t &= \frac{1}{\lambda}(\mathbf{P}_{t-1} - \mathbf{k}_t \mathbf{z}_t^\top \mathbf{P}_{t-1}) \label{eq:rls_cov}
\end{align}
Equation~\eqref{eq:rls_gain} computes a gain vector $\mathbf{k}_t$ that determines how much to adjust the weights.
Equation~\eqref{eq:rls_weight} updates the weight vector proportionally to the prediction error $(y_t - \mathbf{w}_{t-1}^\top \mathbf{z}_t)$.
Equation~\eqref{eq:rls_cov} updates the inverse covariance matrix $\mathbf{P}_t$ to reflect the new observation.
The forgetting factor $\lambda \in (0, 1]$ controls how quickly the model discounts older data: setting $\lambda{<}1$ exponentially downweights past observations, with an effective memory window of approximately $1/(1{-}\lambda)$ samples (e.g., $\lambda{=}0.999$ corresponds to roughly 1{,}000 recent samples, or about 3.5~days at 5-minute intervals).

We warm-start $\mathbf{w}_0$ from the batch Ridge solution and $\mathbf{P}_0$ from $(\tilde{\mathbf{Z}}^\top \tilde{\mathbf{Z}} + \alpha \mathbf{I})^{-1}$, so the online phase begins with an already well-fitted model rather than learning from scratch.
Each RLS update requires $\mathcal{O}(d^2)$ operations, which is negligible for $d{=}36$.
The forgetting factor $\lambda$ is the only hyperparameter, selected from $\{0.998, 0.999, 0.9995\}$ on the validation set.

\subsection{Distributed Learning and Edge Deployment}

A key advantage of per-sensor Ridge regression is that it enables \emph{fully distributed} training and inference.
Since each sensor's model depends only on its own historical flow values, there is no need to collect data from neighboring sensors or a central server.
Each sensor can independently construct its feature vector, solve the Ridge regression in closed form, and perform RLS updates --- all using only locally available data.

This stands in contrast to GCN-based methods, which require a global adjacency matrix and joint optimization across all $N$~sensors, and even to GLMST, whose shared weight matrices ($\mathbf{W}_{t1}$, $\mathbf{W}_{t2}$, $\mathbf{W}_{s1}$) must be trained centrally on data from all sensors.
These methods impose a centralized data collection and training pipeline that may be impractical in large-scale ITS deployments.

Our approach requires only 444~parameters and a $37{\times}37$ matrix inversion per sensor, making it feasible to train and run on edge devices such as roadside units or embedded controllers at individual sensor locations.
RLS online adaptation further eliminates the need for periodic retraining: the model continuously adapts to local traffic pattern changes without communicating with a central system.

To validate this claim, we implement the complete RLS Ridge pipeline in C on an ESP32 microcontroller (Xtensa LX6, 160~MHz, 520~KB SRAM, 4~MB flash), representative of hardware deployed at roadside traffic sensors.
For the embedded implementation, we use a leaner feature set of $d{=}25$ features per sample: 12~recent lags, 1~daily-aligned lag, and 12~weekly-aligned lags, which achieves 15.74\% MAPE on PEMS03 with only 312~parameters per sensor (Table~\ref{tab:complexity}).
The 11 dropped daily lags reduce model size and per-update cost on the ESP32; the accuracy gap to the full $d{=}36$ variant is approximately 0.2--0.3 percentage points on PEMS03 (Table~\ref{tab:complexity} reports both configurations), and the leaner variant still outperforms GLMST.

\textbf{Deployment lifecycle.}
An edge device operating at a single sensor progresses through two phases.
In the \emph{cold-start} phase, the device accumulates raw flow measurements until it has enough history to build an initial Ridge model --- at minimum two weeks to support the weekly-aligned features ($2016 + 11 = 2027$ lookback steps), though more data generally improves the initial model.
In the \emph{online} phase, the device switches to incremental RLS updates as each new measurement arrives, and can periodically flush old data, retaining only the lookback window needed for feature construction (2027~timesteps ${\times}$~4~bytes ${\approx}$~8~KB for one sensor).

As a concrete example, our ESP32 benchmark stores 6~weeks of raw flow for one PEMS08 sensor (17{,}856 timesteps, 70~KB) and uses 4~weeks for initial model training.
Even this generous history fits comfortably in the ESP32's 4~MB flash.
A more constrained deployment using only 2~weeks of training data would require roughly 16~KB of flow storage.

\textbf{Streaming $\mathbf{X}^\top\mathbf{X}$ accumulation.}
During cold-start training, the full feature matrix $\mathbf{X} \in \mathbb{R}^{n \times d}$ would require $n \times d \times 4$~bytes of heap, over 1.2~MB for $n{=}12{,}257$ training samples, exceeding the ESP32's available RAM.
We instead accumulate $\mathbf{X}^\top\mathbf{X}$ and $\mathbf{X}^\top\mathbf{y}$ one row at a time by computing each feature vector on-the-fly from the stored flow via indexed lookups.
Recall from Eq.~\eqref{eq:ridge_closed} that the closed-form solution is $\tilde{\mathbf{w}} = (\tilde{\mathbf{Z}}^\top \tilde{\mathbf{Z}} + \alpha \mathbf{I})^{-1} \tilde{\mathbf{Z}}^\top \mathbf{y}$; rather than inverting the $(d{+}1){\times}(d{+}1)$ Gram matrix $\mathbf{G} = \tilde{\mathbf{Z}}^\top \tilde{\mathbf{Z}}$ (the matrix of all pairwise inner products between feature columns) directly, we compute its Cholesky factorization $\tilde{\mathbf{Z}}^\top \tilde{\mathbf{Z}} + \alpha \mathbf{I} = \mathbf{L}\mathbf{L}^\top$, where $\mathbf{L}$ is lower-triangular, and recover the weight vector $\tilde{\mathbf{w}}$ by forward substitution ($\mathbf{L}\mathbf{v} = \tilde{\mathbf{Z}}^\top\mathbf{y}$) followed by back substitution ($\mathbf{L}^\top\tilde{\mathbf{w}} = \mathbf{v}$).
This avoids an explicit matrix inverse while remaining numerically stable on fixed-point-free hardware.
The entire pipeline, cold-start batch solve, RLS walk-forward, and sliding-window refit, runs with \emph{zero heap allocation}, using approximately 5~KB of stack.

\textbf{Incremental RLS update.}
Once the initial model is built, each new observation triggers the Sherman--Morrison update (Eqs.~\ref{eq:rls_gain}--\ref{eq:rls_cov}): a $d{\times}d$ matrix-vector product, a dot product, and a rank-1 outer-product update.
For $d{=}25$, this completes in 161~$\mu$s per horizon on the ESP32 using single-precision float32 arithmetic.
Predictions match the Python float64 reference implementation to four decimal places, confirming that float32 precision is sufficient for this model size.
Section~\ref{sec:edge_bench} reports detailed timing and resource benchmarks on both the ESP32 and a Raspberry Pi~5.

    \section{Experiments}
\label{sec:experiments}

\subsection{Experimental Setup}

\textbf{Datasets.}
We evaluate on four widely-used PEMS traffic flow benchmarks~\cite{chen2001freeway}, summarized in Table~\ref{tab:datasets}.
All datasets use 5-minute aggregation intervals.
PEMS04 and PEMS08 include speed and occupancy features in addition to flow, though our method uses only the flow channel.
Preliminary experiments incorporating speed and occupancy as additional input features did not improve accuracy; we therefore present flow-only results throughout for simplicity.
Following prior work~\cite{zhang2025glmst,choi2022stgncde}, we split each dataset chronologically in a 6:2:2 ratio for training, validation, and testing.

\textbf{Metrics.}
We report Mean Absolute Error (MAE), Root Mean Squared Error (RMSE), and Mean Absolute Percentage Error (MAPE), averaged over all 12 forecast horizons (5--60 minutes).
MAPE is computed only on timesteps where the actual flow exceeds 5~vehicles, as near-zero flows produce unreliable percentage errors.

\textbf{Baselines.}
We compare against classical methods (VAR~\cite{lutkepohl2005var}, FC-LSTM~\cite{hochreiter1997lstm}), GCN-based methods (STGCN~\cite{yu2018stgcn}, DCRNN~\cite{li2018dcrnn}, Graph WaveNet~\cite{wu2019graphwavenet}, ASTGCN(r)~\cite{guo2019astgcn}, AGCRN~\cite{bai2020agcrn}, STFGNN~\cite{li2021stfgnn}, STG-NCDE~\cite{choi2022stgncde}), and the graph-less MLP baseline GLMST~\cite{zhang2025glmst}.
Baseline numbers are taken directly from~\cite{zhang2025glmst}.

\textbf{Implementation details.}
The PEMS datasets contain no NaN entries (already preprocessed by the data providers); zero values are retained as observations, and the MAPE filter at 5~vehicles (above) excludes near-zero targets from the percentage-error metric, matching the protocol used by all reported baselines.
Ridge regression uses regularization $\alpha{=}1.0$.
The feature vector for each sensor and horizon~$h$ consists of 12~recent flow values, 12~daily-aligned values, and 12~weekly-aligned values ($d{=}36$).
Each Ridge model has $d{+}1{=}37$ parameters per horizon, giving $37{\times}12{=}444$ total parameters per sensor.
For RLS, we warm-start from the Ridge solution and perform online updates on the test set with forgetting factor $\lambda$ selected from $\{0.998, 0.999, 0.9995\}$ by validation MAPE.
At each test step $t$, the per-horizon RLS model predicts $\hat{x}_i(t{+}h)$ from current weights and (for $t \geq h$) updates using $\mathbf{z}_i^{(h)}(t{-}h)$ paired with the now-observable target $x_i(t)$, so updates lag predictions by $h$ steps.
The pipeline is deterministic (Ridge has a closed-form solution and RLS is a deterministic Sherman--Morrison sequence), so we do not report run-to-run variance.
All experiments run on a single CPU core.

\begin{table}[!t]
\centering
\caption{Dataset statistics. F = flow, S = speed, O = occupancy.}
\label{tab:datasets}
\renewcommand{\arraystretch}{1.1}
\begin{tabular}{lcccc}
\toprule
Dataset & Sensors & Timesteps & Days & Features \\
\midrule
PEMS03 & 358 & 26{,}208 & 91 & F \\
PEMS04 & 307 & 16{,}992 & 59 & F, S, O \\
PEMS07 & 883 & 28{,}224 & 98 & F \\
PEMS08 & 170 & 17{,}856 & 62 & F, S, O \\
\bottomrule
\end{tabular}
\end{table}

\begin{table*}[t]
\centering
\caption{Performance comparison on four PEMS benchmarks. Best result in \textbf{bold}, second best \underline{underlined}. Baseline numbers from~\cite{zhang2025glmst}. Bottom row shows RLS Ridge improvement over GLMST (negative = better).}
\label{tab:main}
\setlength{\tabcolsep}{3.5pt}
\begin{tabular}{l@{\hskip 8pt}ccc@{\hskip 8pt}ccc@{\hskip 8pt}ccc@{\hskip 8pt}ccc}
\toprule
& \multicolumn{3}{c}{PEMS03} & \multicolumn{3}{c}{PEMS04} & \multicolumn{3}{c}{PEMS07} & \multicolumn{3}{c}{PEMS08} \\
\cmidrule(lr){2-4} \cmidrule(lr){5-7} \cmidrule(lr){8-10} \cmidrule(lr){11-13}
Model & MAE & RMSE & MAPE & MAE & RMSE & MAPE & MAE & RMSE & MAPE & MAE & RMSE & MAPE \\
\midrule
\multicolumn{13}{l}{\textit{Classical}} \\
VAR & 23.65 & 38.26 & 24.51 & 24.54 & 38.61 & 17.24 & 50.22 & 75.63 & 32.22 & 19.19 & 29.81 & 13.10 \\
FC-LSTM & 21.33 & 35.11 & 23.33 & 26.77 & 40.65 & 18.23 & 29.98 & 45.94 & 13.20 & 23.09 & 35.17 & 14.99 \\
\midrule
\multicolumn{13}{l}{\textit{GCN-based}} \\
STGCN & 17.55 & 30.42 & 17.34 & 21.16 & 34.89 & 13.83 & 25.33 & 39.34 & 11.21 & 17.50 & 27.09 & 11.29 \\
DCRNN & 17.99 & 30.31 & 18.34 & 21.22 & 33.44 & 14.17 & 25.22 & 38.61 & 11.82 & 16.82 & 26.36 & 10.92 \\
GraphWaveNet & 19.12 & 32.77 & 18.89 & 24.89 & 39.66 & 17.29 & 26.39 & 41.50 & 11.97 & 18.28 & 30.05 & 12.15 \\
ASTGCN(r) & 17.34 & 29.56 & 17.21 & 22.93 & 35.22 & 16.56 & 24.01 & 37.87 & 10.73 & 18.25 & 28.06 & 11.64 \\
AGCRN & 15.98 & 28.25 & 15.23 & 19.83 & 32.26 & 12.97 & 22.37 & 36.55 & 9.12 & 15.95 & 25.22 & 10.09 \\
STFGNN & 16.77 & 28.34 & 16.30 & 20.48 & 32.51 & 16.77 & 23.46 & 36.60 & 9.21 & 16.94 & 26.25 & 10.60 \\
STG-NCDE & \underline{15.57} & 27.09 & 15.06 & \underline{19.21} & 31.09 & \underline{12.76} & \textbf{20.53} & 33.84 & 8.80 & 15.45 & 24.81 & 9.92 \\
\midrule
\multicolumn{13}{l}{\textit{Graph-less MLP}} \\
GLMST & \textbf{15.21} & \underline{26.77} & \underline{14.74} & \textbf{19.00} & 31.20 & \textbf{12.46} & \underline{20.73} & 34.60 & \underline{8.76} & 15.48 & 24.69 & 9.78 \\
GLMST(4, 32, 4) & 18.57 & 30.06 & 15.50 & 22.92 & 35.96 & 13.90 & 25.57 & 39.59 & 10.60 & 18.33 & 28.64 & 10.39 \\
\midrule
\multicolumn{13}{l}{\textit{Linear (Ours)}} \\
Static Ridge & 18.00 & 26.77 & 15.47 & 21.04 & \underline{30.12} & 13.18 & 21.99 & \underline{31.00} & 9.04 & \underline{15.22} & \underline{21.34} & \underline{9.52} \\
RLS Ridge & 16.19 & \textbf{23.47} & \textbf{13.78} & 20.86 & \textbf{29.91} & 13.04 & 20.86 & \textbf{29.53} & \textbf{8.50} & \textbf{14.63} & \textbf{20.64} & \textbf{8.82} \\
\midrule
RLS vs.\ GLMST & +0.98 & $-$3.30 & $-$0.96 & +1.86 & $-$1.29 & +0.58 & +0.13 & $-$5.07 & $-$0.26 & $-$0.85 & $-$4.05 & $-$0.96 \\
\bottomrule
\end{tabular}
\end{table*}

\subsection{Main Results}

Table~\ref{tab:main} presents the main comparison.
We report both Static Ridge (frozen weights at test time) and RLS Ridge (online-adaptive); all baseline numbers are reproduced from~\cite{zhang2025glmst} with frozen weights, so RLS Ridge is the only entry that adapts during evaluation.
Our RLS Ridge achieves the best MAPE on three of four datasets (PEMS03, PEMS07, PEMS08) while using only 444~parameters per sensor, an $80{\times}$ reduction compared to GLMST's ${\sim}$35{,}500 per-sensor footprint (35K shared $+$ 512 node-specific).
On the remaining dataset (PEMS04), the gap to the best baseline is less than one percentage point (13.04\% vs.\ GLMST's 12.46\%).

On PEMS03, RLS Ridge achieves 13.78\% MAPE, improving over GLMST's 14.74\%.
On PEMS07 and PEMS08, RLS Ridge achieves 8.50\% and 8.82\%, surpassing all baselines including GCN methods.
The static Ridge baseline already matches or exceeds most GCN methods on PEMS07 (9.04\% vs.\ AGCRN's 9.12\%) and PEMS08 (9.52\% vs.\ STFGNN's 10.60\%).
The compressed variant GLMST(4, 32, 4), which reduces the embedding dimension from 64 to 4 (see Section~\ref{sec:nas}), degrades MAPE by 0.8--1.8 percentage points across all datasets, confirming that the full GLMST's capacity is largely redundant.

\subsection{Complexity Analysis}
\label{sec:nas}

Table~\ref{tab:complexity} traces the complexity spectrum on PEMS03, from the original GLMST down to our Ridge models.
To enable a fair comparison, we report \emph{per-sensor} parameter counts: for each GLMST variant, this includes both the shared parameters (which every sensor requires for inference) and the node-specific parameters $\mathbf{w}_{s2}^{(n)}, \mathbf{b}_{s2}^{(n)}$ for that sensor.
Ridge models have no shared parameters; each sensor's model is self-contained.

We instantiate GLMST with $e{=}64$, $s{=}64$, $L{=}4$ following our reproduction, yielding 35{,}585~per-sensor parameters (35K shared across all sensors $+$ 512 node-specific; 218K total for $N{=}358$; see Section~\ref{sec:glmst_bottleneck}).
A grid search over the GLMST($e$, $s$, $L$) configurations reveals that aggressive compression preserves most accuracy: GLMST(4, 24, 4) achieves 15.42\% MAPE with 2{,}157~per-sensor parameters, and even GLMST(4, 8, 2) reaches 15.73\% with under 1K.
At the extreme, GLMST(1, 8, 2), where temporal information flows through a single scalar, still achieves 16.77\%, only 2 points behind the full model.

Replacing the neural architecture entirely with Ridge regression using periodic features yields 15.47\% MAPE with 444~parameters per sensor (12r+12d+12w).
A leaner feature set (12r+1d+12w) achieves 15.74\% with only 312~parameters.
Adding RLS online adaptation improves both variants: the leaner set (12r+1d+12w) reaches 14.02\% and the full set (12r+12d+12w) achieves 13.78\%, the best result on PEMS03.
Unlike GLMST, where even the compressed variants require centralized training on all sensors' data and subsequent distribution of the shared weights, Ridge and RLS Ridge train entirely from local data at each sensor.

\begin{table}[t]
\centering
\caption{Complexity spectrum on PEMS03 ($N{=}358$). Per-sensor parameters include shared $+$ node-specific for GLMST variants. Ridge models have no shared parameters. GLMST totals scale with $N$ (122K--487K across PEMS datasets). Feature notation: 12r = 12 recent lags, 12d/1d = 12/1 daily-aligned, 12w = 12 weekly-aligned.}
\label{tab:complexity}
\setlength{\tabcolsep}{4pt}
\begin{tabular}{llrc}
\toprule
Model & Config & Per-sensor & MAPE \\
\midrule
GLMST~\cite{zhang2025glmst} & $e{=}64$, $s{=}64$, $L{=}4$ & 35{,}585 & 14.74 \\
\midrule
GLMST(4, 24, 4) & $e{=}4$, $s{=}24$, $L{=}4$ & 2{,}157 & 15.42 \\
GLMST(4, 8, 3) & $e{=}4$, $s{=}8$, $L{=}3$ & 1{,}201 & 15.49 \\
GLMST(4, 8, 2) & $e{=}4$, $s{=}8$, $L{=}2$ & 949 & 15.73 \\
GLMST(2, 8, 2) & $e{=}2$, $s{=}8$, $L{=}2$ & 859 & 15.94 \\
GLMST(1, 8, 2) & $e{=}1$, $s{=}8$, $L{=}2$ & 814 & 16.77 \\
\midrule
Static Ridge & 12r only & 156 & 20.14 \\
Static Ridge & 12r + 1d + 12w & 312 & 15.74 \\
Static Ridge & 12r + 12d + 12w & 444 & 15.47 \\
RLS Ridge & 12r + 1d + 12w, $\lambda{=}0.998$ & 312 & 14.02 \\
RLS Ridge & 12r + 12d + 12w, $\lambda{=}0.998$ & 444 & \textbf{13.78} \\
\bottomrule
\end{tabular}
\end{table}

\subsection{Ablation Study}

\textbf{Feature importance.}
Table~\ref{tab:complexity} reveals the contribution of each feature group.
Using only 12~recent timesteps yields 20.14\% MAPE on PEMS03.
Adding horizon-aligned daily and weekly lags reduces MAPE to 15.47\% ($-$4.67~pp), confirming that periodic patterns are the dominant predictive signal.
Online RLS adaptation further reduces MAPE from 15.47\% to 13.78\% ($-$1.69~pp), capturing distribution shift during the test period.

\begin{figure}[t]
\centering
\includegraphics[width=\columnwidth]{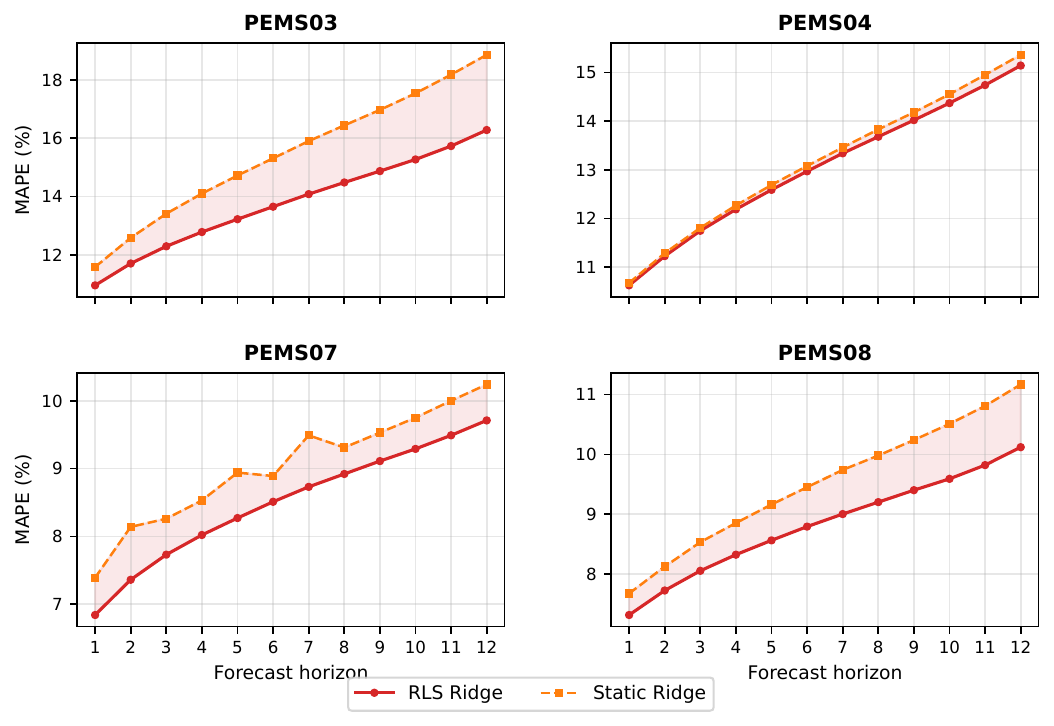}
\caption{Per-horizon MAPE across four datasets. RLS Ridge (solid red) consistently improves over static Ridge (dashed orange), with the shaded region showing the gain from online adaptation.}
\label{fig:horizon}
\end{figure}

\textbf{Per-horizon analysis.}
Fig.~\ref{fig:horizon} shows MAPE as a function of forecast horizon.
RLS Ridge consistently outperforms static Ridge at every horizon, with the gap widening at longer horizons where adaptation to recent patterns matters most.

\begin{table}[t]
\centering
\caption{Computational cost comparison on PEMS07 (883~sensors, $d{=}36$ features). Static and RLS Ridge are per-sensor on a single CPU core; GLMST variants train jointly on a V100 GPU.}
\label{tab:efficiency}
\setlength{\tabcolsep}{3pt}
\begin{tabular}{lcccc}
\toprule
& GLMST & GLMST$^\ddagger$ & Static Ridge & RLS Ridge \\
\midrule
Training hardware & GPU & GPU & CPU & CPU \\
Training scope & joint & joint & per sensor & per sensor \\
Training time & 78 min & 37 min & 3 min & 3 min \\
Training per sensor & --- & --- & 0.17 s & 0.17 s \\
\midrule
Inference FLOPs/step$^\dagger$ & 974K & 46K & 876 & 876 \\
Update FLOPs/step$^\dagger$ & --- & --- & --- & 81K \\
\bottomrule
\multicolumn{5}{l}{\footnotesize $^\dagger$Per sensor, all 12 horizons, one timestep.} \\
\multicolumn{5}{l}{\footnotesize $^\ddagger$Compressed: $e{=}4$, $s{=}32$, $L{=}4$.}
\end{tabular}
\end{table}

\textbf{Spatial features.}
We experimented with incorporating neighboring-sensor flow values as additional features.
This consistently degraded performance, likely because Ridge regression cannot learn the non-linear spatial interactions that GCN layers capture.
We therefore omit spatial features in our final model.

\subsection{Efficiency}

Table~\ref{tab:efficiency} compares computational cost across three dimensions: training, inference, and online adaptation.

\textbf{Training.}
GLMST trains all sensors jointly via backpropagation on a GPU.
On PEMS07 (883~sensors), the full model ($e{=}64$, $s{=}64$, $L{=}4$) takes ${\sim}$78~minutes (100~epochs $\times$ 47~s/epoch) on a V100; the compressed GLMST(4,32,4) variant trains in ${\sim}$37~minutes (22~s/epoch).
Ridge regression solves each sensor independently in closed form: forming $\mathbf{X}^\top\mathbf{X}$ and solving the $36{\times}36$ linear system takes ${\sim}$40M~FLOPs per horizon (${\sim}$477M for all 12~horizons), or 0.17~s per sensor on a single CPU core (Xeon E5-2640 v4, 2.4~GHz).
The entire PEMS07 Ridge pipeline completes in under 3~minutes on one core, over $10{\times}$ faster than even the compressed GLMST on a GPU, with no inter-sensor communication.

\textbf{Inference and online update.}
GLMST inference involves input projection, temporal fusion, and $L$~spatio-temporal blocks with shared linear layers, totaling ${\sim}$974K~FLOPs per sensor for the full model ($e{=}64$, $s{=}64$, $L{=}4$).
By contrast, Ridge inference requires only a single dot product per horizon: 876~FLOPs for all 12~horizons, over $1{,}000{\times}$ cheaper.
The RLS predict-and-update cycle adds a rank-1 matrix update per horizon: the dominant cost is maintaining the $d{\times}d$ inverse covariance matrix~$\mathbf{P}$ via the Sherman--Morrison update ($5d^2$ FLOPs per horizon), totaling ${\sim}$81K~FLOPs across all 12~horizons per sensor per timestep. This is still $12{\times}$ cheaper than GLMST inference alone, and feasible even on a microcontroller (Section~\ref{sec:edge_bench}).

\subsection{Edge Deployment Benchmarks}
\label{sec:edge_bench}

To demonstrate practical deployability, we benchmark the RLS Ridge pipeline on an ESP32 microcontroller and a Raspberry Pi~5.
All benchmarks use PEMS08 with the 25-feature configuration and 12{,}257~training samples (${\sim}$6~weeks of 5-minute data).

\textbf{Per-sensor timing.}
Table~\ref{tab:edge_timing} compares per-sensor training and inference times on both platforms.
On the ESP32 (Xtensa LX6, 160~MHz, float32 C), cold-start training (streaming $\mathbf{X}^\top\mathbf{X}$ accumulation over 12{,}257~samples followed by Cholesky solve) completes in 7.4~s for all 12~horizons.
A single predict-and-update cycle then takes 1.93~ms for all 12~horizons, representing $<$0.001\% of the 5-minute measurement interval.
On the Raspberry Pi~5 (Cortex-A76, 2.4~GHz, float64 Python), training takes 0.21~s per sensor and predict-and-update takes 0.26~ms.
With 4~cores, the Pi~5 handles all 170~PEMS08 sensors in 17.5~ms per step, suitable for corridor-scale deployment.

\begin{table}[t]
\centering
\caption{Per-sensor RLS Ridge timing on edge platforms (PEMS08, 25~features, 12{,}257 training samples $\approx$ 6~weeks). Pi~5 is a 4-core device; per-sensor numbers are single-thread, and the 17.5~ms/step aggregate reported in the text uses all 4 cores.}
\label{tab:edge_timing}
\setlength{\tabcolsep}{4pt}
\begin{tabular}{lcc}
\toprule
& ESP32 & Pi~5 \\
\midrule
Processor & Xtensa LX6 & Cortex-A76 \\
Clock & 160 MHz & 2.4 GHz \\
Precision & float32 (C) & float64 (Python) \\
\midrule
Training (per sensor, 12h) & 7.4 s & 0.21 s \\
Predict+update (per sensor, 12h) & 1.93 ms & 0.26 ms \\
\bottomrule
\end{tabular}
\end{table}

\textbf{ESP32 resource usage.}
Table~\ref{tab:esp32_resources} summarizes the ESP32 memory footprint.
The entire application (code, raw flow data, model state, and standardization parameters) occupies 285~KB of flash and requires only 5~KB of peak stack with zero heap allocation.
This leaves over 380~KB of RAM for the RTOS, WiFi/BLE communication, and other sensor tasks.

\begin{table}[t]
\centering
\caption{ESP32 resource usage for single-sensor RLS Ridge (PEMS08, 25 features, 12 horizons).}
\label{tab:esp32_resources}
\setlength{\tabcolsep}{4pt}
\begin{tabular}{lrl}
\toprule
Component & Size & Notes \\
\midrule
Raw flow timeseries & 69.8 KB & 17{,}856 float32 values \\
Model state ($\mathbf{w}$, $\mathbf{P}$, $\alpha$) & 30.5 KB & 12 horizons \\
Standardization params & 2.4 KB & $\mu_x$, $\sigma_x$, $\mu_y$ \\
Compiled binary (total) & 285 KB & 7\% of 4 MB flash \\
\midrule
Peak stack (runtime) & 5 KB & Zero heap allocation \\
\bottomrule
\end{tabular}
\end{table}

\textbf{Bounded storage.}
In continuous operation, the device need not retain the full flow history indefinitely.
Because RLS maintains the model state in the inverse covariance matrix $\mathbf{P}$ and weight vector $\mathbf{w}$, raw observations older than a sliding window (e.g., 6~weeks) can be discarded without retraining.
The forgetting factor $\lambda$ naturally down-weights older data, so the model adapts to recent patterns while the stored flow footprint remains bounded.

	\section{Summary and Conclusions}
\label{sec:conclusion}

We investigated how much model complexity is actually needed for short-term traffic flow prediction.
A parametric analysis of GLMST revealed that, because each sensor observes a single scalar flow value per timestep, the hidden representations produced by the input projection are inherently rank-1.
Consequently, the temporal mixer's effective capacity collapses to at most four dimensions, and a 64-dimensional embedding adds parameters without extracting additional information.
This motivated replacing the neural architecture with per-sensor Ridge regression using horizon-aligned periodic features.
Combined with RLS online adaptation, this approach achieves the best MAPE on three of four PEMS benchmarks while using only 444~parameters per sensor, an $80{\times}$ reduction compared to GLMST's per-sensor footprint.
We validated fully distributed edge deployment: an ESP32 completes inference and model updates for all 12~horizons in under 2~ms with zero heap allocation, and a Raspberry Pi~5 handles 170~sensors in 17.5~ms per step.

Future work includes characterizing the fundamental accuracy limits of edge-deployable per-sensor models and exploring multi-sensor coordination where neighboring sensors share compact model summaries rather than raw data.

	\balance
	\bibliographystyle{IEEEtran}
	\bibliography{root}

\end{document}